\documentclass[letterpaper,11pt,fleqn]{article}
\usepackage{jheppub}
\usepackage{bm,mathrsfs,slashed}

\long\def\comment#1{ }
\newcommand{\eqnum}[1]{eq.~\eqref{#1}}
\newcommand{\eqn}[1]{eq.~\eqref{#1}}

\newcommand{\beq}{\begin{eqnarray}}
\newcommand{\eeq}{\end{eqnarray}}
\newcommand{\nn}{\nonumber\\}

\newcommand{\dif}{{\rm d}}

\newcommand{\rmd}{{\rm d}}

\newcommand{\EA}{\mcal{E}_{\rm A}}
\newcommand{\EB}{\mcal{E}_{\rm B}}
\newcommand{\Eone}{\mcal{E}_{\rm rad}^{(1)}}
\newcommand{\Etwo}{\mcal{E}_{\rm rad}^{(2)}}

\newcommand{\mW}{\mcal{W}}

\newcommand{\del}{\partial}

\newcommand{\order}[1]{\mcal{O}{(#1)}}
\newcommand{\mcal}{\mathcal}

\newcommand{\tp}{\acute{t}}
\newcommand{\xp}{\acute{x}}

\newcommand{\rp}{\acute{r}}
\newcommand{\tti}{\tilde{t}}
\newcommand{\bxT}{\bm{x}_{\perp}}
\newcommand{\br}{\bm{r}}
\newcommand{\brp}{\acute{\bm{r}}}

\newcommand{\brs}{\bm{r}_s}
\newcommand{\dbrs}{\dot{\bm{r}}_s}
\newcommand{\pbrs}{\bm{r}_s'}
\newcommand{\brq}{\bm{r}_q}
\newcommand{\bvq}{\bm{\upsilon}_q}
\newcommand{\baq}{\bm{a}_q}
\newcommand{\dbaq}{\dot{\bm{a}}_q}
\newcommand{\vq}{\upsilon_q}
\newcommand{\gq}{\gamma_q}

\let\Oldcdot\cdot
\renewcommand{\cdot}{\mspace{-2mu}\Oldcdot\mspace{-2mu}}
\let\Oldtimes\times
\renewcommand{\times}{\mspace{-2mu}\Oldtimes\mspace{-2mu}}

\title{\Large Radiation by a heavy quark in $\mcal{N}=4$ SYM at strong coupling}

\author[a]{Y.~Hatta,}
\author[b,c]{E.~Iancu,}
\author[d]{A.H.~Mueller,}
\author[e]{and D.N.~Triantafyllopoulos}

\affiliation[a]{Graduate School of Pure and Applied Sciences,
University of Tsukuba,\\ Tsukuba, Ibaraki 305-8571, Japan}
\affiliation[b]{Institut de Physique Th\'{e}orique de Saclay,
F-91191 Gif-sur-Yvette, France}
\affiliation[c]{CERN, Theory Division, CH-1211 Geneva, Switzerland}
\affiliation[d]{Department of Physics, Columbia University, New York, NY
10027, U.S.A.}
\affiliation[e]{ECT*, European Center for Theoretical Studies in Nuclear
Physics and Related Areas,\\ Strada delle Tabarelle 286, I-38123 Villazzano
(TN), Italy}

\emailAdd{hatta@het.ph.tsukuba.ac.jp}
\emailAdd{edmond.iancu@cea.fr}
\emailAdd{amh@phys.columbia.edu}
\emailAdd{trianta@ect.it}

\abstract{Using the AdS/CFT correspondence in the supergravity
approximation, we compute the energy density radiated by a heavy quark
undergoing some arbitrary motion in the vacuum of the strongly coupled
${\mathcal N}=4$ supersymmetric Yang--Mills theory. We find that this
energy is fully generated via backreaction from the near--boundary endpoint
of the dual string attached to the heavy quark. Because of that, the energy
distribution shows the same space--time localization as the classical
radiation that would be produced by the heavy quark at weak coupling.
We believe that this and some other unnatural features of our result
(like its anisotropy and the presence of regions with negative energy
density) are artifacts of the supergravity approximation, which will be
corrected after including string fluctuations.
For the case where the quark trajectory is bounded, we also compute the
radiated power, by integrating the energy density over the surface of a
sphere at infinity. For sufficiently large times, we find agreement with
a previous calculation by Mikhailov {\tt [hep-th/0305196]}.
}
\begin{document}
\maketitle

\section{Introduction}
\label{intro}

One of the basic problems that one can think of in the context of any
gauge theory, and in particular within a strongly--coupled, conformal,
field theory as described by the AdS/CFT correspondence
\cite{Maldacena:1997re,Gubser:1998bc,Witten:1998qj}, is that of the
radiation by a moving, classical, charged particle. By `classical' we
mean a particle which is heavy enough to be treated as pointlike and
assumed to follow a well--identified, classical, trajectory (say, under
the action of an external force). And by `radiation' we mean the emission
of quanta of the underlying gauge theory which escape at infinity, thus
generating energy loss. For asymptotically weak coupling, these quanta
need to be strictly on--shell and hence propagate at the speed of light
(for the radiation in the vacuum). But in general, the emitted quanta can
be also off--shell, in which case they are subjected to further evolution
(e.g., time--like quanta can decay). In particular, when the coupling is
strong, we expect such off--shell effects to be very important and
generate a very different space--time pattern for the radiated energy as
compared to weak coupling.

Consider, for instance, the radiation produced by a heavy quark subjected
to a kick, {\em i.e.} an external force which is localized in space and
time. In a classical calculation, which is the same as the weak coupling
limit of the corresponding field theory, the radiation will propagate
away from the quark as a spherical shell which is radially expanding at
the speed of light ($r=t$), with a small width $\Delta r$ determined by
the duration of the original perturbation. This is quite different from
the picture we would expect at strong coupling \cite{HIM3}. There, the
radiation should typically proceed via the emission of a few virtual
quanta, which will then decay into other quanta, thus eventually
generating a system of partons with a wide distribution in virtualities.
Since time--like quanta propagate slower than the speed of light, the
energy taken away by those quanta should exhibit a rather broad
distribution in $r$ at $r\lesssim t$. Since moreover the various quanta
can be randomly emitted along any direction (in the quark rest frame),
this picture also implies that the energy distribution should be
isotropic (up to a Lorentz boost). This last prediction has been checked
via an explicit calculation within AdS/CFT of the angular distribution of
the energy produced by the decay of a time--like wavepacket at strong
coupling \cite{Hofman:2008ar}.

In view of the above, it appeared as a surprise when other AdS/CFT
calculations \cite{Athanasiou:2010pv,Hatta:2010dz}, which have also
investigated the radial distribution, found that there is no broadening
(at least, within the limits of the respective calculations): the energy
radiated within the vacuum of the ${\mathcal N}=4$ supersymmetric
Yang--Mills (SYM) theory at strong coupling appears to be as sharply
localized in $r$ as the corresponding classical result. This was first
noticed in Ref.~\cite{Athanasiou:2010pv} for the example of the
synchrotron radiation produced by a heavy quark in uniform rotation and
then extended in Ref.~\cite{Hatta:2010dz} to other situations, including
the two problems alluded to above --- a heavy quark perturbed by a kick
and the decay of a time--like `photon'. As pointed out in
\cite{Hatta:2010dz} (but already visible at the level of the calculations
in \cite{Athanasiou:2010pv}), this lack of radial broadening is to be
attributed to the fact that, within the supergravity approximation used
in these calculations, the whole contribution to radiation is generated
via backreaction from points near the Minkowski boundary of AdS$_5$.

At this level, it is useful to recall that the supergravity approximation
is the classical limit of the dual string theory, which neglects string
loops and string fluctuations, and is generally accepted to faithfully
describe the strong `t Hooft coupling limit $\lambda=g^2N_c\to\infty$
with fixed $g\ll 1$ of the ${\mathcal N}=4$ SYM theory
\cite{Maldacena:1997re,Gubser:1998bc,Witten:1998qj}. ($g$ is the
Yang--Mills coupling and $N_c$ the number of colors.) Furthermore, the
`backreaction' refers to the AdS/CFT calculation of the energy density in
the gauge theory, which involves the response of the AdS$_5$ metric to
the 5D stress tensor of the bulk object dual to the physical excitation
on the boundary. (For instance, this bulk object is a Nambu--Goto string
in the case of a heavy quark in the fundamental representation of
SU$(N_c)$, and a supergravity vector field wave--packet in the case of a
time--like photon.)

{\em A priori}, the calculation of the backreaction involves an integral
over all the points within the support of the bulk stress tensor, say
along the string in the case of a heavy quark. Similar calculations at
{\em finite temperature}
\cite{Gubser:2007xz,Chesler:2007an,Chesler:2008wd,Dominguez:2008vd} have
shown that, in that context, all the points along the string provide
non--trivial contributions to the energy density on the boundary, which
therefore shows broadening: points of the string which lie further and
further away from the boundary provide contributions which are more and
more spread in space--time. This is in the spirit of the {\em
ultraviolet/infrared correspondence} \cite{Susskind:1998dq,Peet:1998wn},
which associates increasing distance from the boundary of AdS$_5$ with
increasing virtuality in the original gauge theory. However, in the case
of the radiation in the {\em vacuum} (say, as produced by an accelerated
quark), the calculations in Refs.~\cite{Athanasiou:2010pv,Hatta:2010dz}
show that the integral expressing the backreaction reduces to a boundary
contribution from the string endpoint at the Minkowski boundary. Thus,
there is effectively no virtuality involved in this calculation, which
`explains' why the final result shows no spreading. But this
`explanation' leaves us with a physical paradox, namely why should
radiation in a quantum field theory at infinitely strong coupling involve
only on--shell (light--like) modes, without any trace of virtual quantum
fluctuations.

In Ref.~\cite{Hatta:2010dz} we have also proposed a possible solution to
this puzzle, by identifying a class of stringy corrections which are {\em
not} suppressed in the strong coupling limit and which when included in
the calculation of the backreaction --- in an admittedly heuristic way,
by lack of a proper treatment of string fluctuations in AdS$_5$
--- seem to provide energy broadening, in conformity with the UV/IR
correspondence. It would be of course very interesting to make further
progress with the understanding of string corrections in AdS$_5$, which
is an outstanding open problem. This is however not the purpose of this
paper. Rather, here we shall be more modest and extend the results in
Ref.~\cite{Hatta:2010dz} in a different direction: we shall provide an
exact result for the energy density radiated in the supergravity
approximation by heavy quark undergoing some arbitrary motion in the
vacuum of the ${\mathcal N}=4$ SYM theory.

In spite of our own criticism of the supergravity approximation for the
type of problems at hand, we believe that the present results are
nevertheless interesting for several reasons. First, a precise knowledge
of the classical, supergravity, result is a first and mandatory step in
any effort aiming at including string corrections. Second, by itself,
this classical calculation is rather non--trivial, as it requires an
exact, analytic, solution to the problem of the backreaction. Previously,
such analytic solutions have been given only for particular cases --- the
most non--trivial one being the calculation of the synchrotron radiation
in Ref.~\cite{Athanasiou:2010pv}. Our general results below will allow us
to simply recover such previous results and extend them to an arbitrary
motion. Third, we shall explicitly verify that, also in the general case,
the whole contribution to the backreaction is still coming from the
string endpoint near the boundary; hence, in this approximation, the
radiation propagates at the speed of light, like in a classical field
theory. Fourth, our results exhibit some other surprising features
(besides the lack of radial broadening), which in our opinion reflect the
limitations of the supergravity approximation: the energy density appears
to be anisotropic and also negative in some regions of space--time. The
anisotropy is unnatural at strong coupling, for reasons explained before;
it is even more so in the context of the ${\mathcal N}=4$ SYM theory,
where, as we shall see, already the corresponding {\em classical} result
is fully isotropic\footnote{To better appreciate the non--trivial
character of this property, one should recall that it does not hold for
radiation in classical electrodynamics \cite{Jackson}. See also the
discussion in Sect.~\ref{classical}.} (up to boost effects). As for the
negativity of the energy density, which was already noticed (for the
example of uniform rotation) in \cite{Athanasiou:2010pv}, this is in
principle acceptable in a quantum field theory, but we find it very
unnatural in the context of radiation, for reasons to be discussed in
Sect.~\ref{discussion}.

Our analysis of the backreaction, that we now outline, will build upon
previous constructions in the literature. An essential ingredient in that
sense is the exact solution, due to Mikhailov \cite{Mikhailov:2003er},
for the string profile corresponding to an arbitrary motion of the heavy
quark. This solution is not fully explicit
--- it still depends upon a `retardation time', determined as the
solution of a generally transcendental equation (see Sect.~\ref{profile}
for details)
---, but this is not less explicit than the usual textbook treatment of
radiation in classical electrodynamics \cite{Jackson}, where the results
are written as a function of the `retardation time' (the time of
emission, related to the time and point of measurement by the condition
of propagation at the speed of light). By studying the energy carried by
the accelerated string, Mikhailov has also deduced a formula for the
radiated power, which appears to be similar to Li\'enard formula in
classical electrodynamics. His results have been extended in
Refs.~\cite{Chernicoff:2009re,Chernicoff:2009ff}, where the total energy
of the moving quark (proper energy plus radiation) has been inferred via
a world--sheet analysis.

Furthermore, in computing the backreaction, we shall use the general
formul\ae{} established in Ref.~\cite{Athanasiou:2010pv} which express
the energy density on the boundary as a convolution between the bulk
stress tensor of the string and the bulk--to--boundary propagator. Using
Mikhailov's solution for the string profile, we shall express the bulk
stress tensor in terms of the quark motion on the boundary (in
Sect.~\ref{bulk}), and this will allow us to explicitly perform the
integrals yielding the backreaction (in Sect.~\ref{back}). We shall thus
find that, due to remarkable cancelations between contributions arising
from various components of the stress tensor, the only terms which are
left in the final result for the energy density are boundary
contributions from the string endpoint at the heavy quark. Then, in
Sect.~\ref{rad}, we shall extract the {\em radiative} energy density,
defined as the component of the total energy showing the slowest decay
($\sim 1/R^2$) at large distances. This is the main result of our paper.
By integrating this result over the surface of a sphere at infinity (an
operation which is well defined when the quark trajectory is confined to
a bounded region in space), we shall also compute the radiated power
(still in Sect.~\ref{rad}). We shall thus find the term originally
obtained by Mikhailov \cite{Mikhailov:2003er} and also a second term,
which is however subleading at large times\footnote{Interestingly, this
second term is similar to a piece of the total energy of the accelerated
string which in Refs.~\cite{Chernicoff:2009re,Chernicoff:2009ff} has been
interpreted as a part of the quark proper (or kinetic) energy. See the
discussion in Sect.~\ref{applications}.}. In the remaining part of the
paper, we shall further discuss our results, compare them to some known
limits in the literature (in Sect.~\ref{applications}) and also to the
corresponding classical results (that we shall derive in the context of
the ${\mathcal N}=4$ SYM theory in Sect.~\ref{classical}). In our final
discussion in Sect.~\ref{discussion}, we shall emphasize some peculiar
features of these results, which point towards limitations of the
supergravity approximation.

\section{The string profile}
\label{profile}

We consider a heavy quark in the fundamental representation of the gauge
group SU$(N_c)$ which undergoes some arbitrary motion, with trajectory
$\br= \brq(t)$, within the vacuum of the ${\mathcal N}=4$ SYM theory at
strong coupling. (We assume the quark to be arbitrarily heavy, so that
the notion of classical trajectory makes indeed sense for it.) The quark
can either be in isolation, or it can be a part of a quark--antiquark
pair. The dual supergravity description of the quark dynamics is a
Nambu--Goto string hanging in AdS$_5$, with one endpoint attached to a
D7--brane. The string can be either finite, with the other endpoint
attached to the same D7--brane (the case of a quark--antiquark pair), or
it can extend all the way to the center of AdS$_5$ (the case of a single
quark).

We shall parameterize the AdS$_5$ space--time using Poincar{\'e}
coordinates, with metric
 \beq\label{metric}
 \rmd s^2\,\equiv\,G_{MN}\,\rmd x^M \rmd x^N
 \,= \,
 \frac{L^2}{z^2} \,\left( -\rmd t^2 + \rmd \br^2 + \rmd z^2 \right).
 \eeq
Here, $x^M=(x^\mu, z)$ where $x^\mu=(t,\br)$ are the Minkowski
coordinates in the physical space--time and $z$ (with $0\le z < \infty$)
is the fifth dimension, also known as the `radial coordinate in AdS$_5$'
(not to be confused with the physical radius $r=|\br|$). In these
coordinates, the Minkowski boundary lies at $z=0$, the center of AdS$_5$
is at $z\to\infty$, and the D7--brane ends at a distance
$z_m=\sqrt{\lambda}/(2\pi m_q)$ from the boundary, with $m_q$ the quark
mass. In what follows we shall choose $m_q$ to be large enough for $z_m$
to be much smaller than any other interesting space--time scale.

The string dynamics is encoded in the Nambu--Goto action
 \beq\label{NG}
 S = -T_0 \int \dif \tau \, \dif \sigma \sqrt{-g}\,,
 \qquad g_{ab}\,=\,G_{MN}\del_a X^M \del_b X^N\,,
 %\sqrt{-g} &=& |G_{00}|\, \sqrt{1-\dbrs^2+\pbrs^2 - (\dbrs \times \pbrs)^2}.
 \eeq
where $T_0 = \sqrt{\lambda}/2\pi L^2$ is the string tension, $\tau$ and
$\sigma$ are the two coordinates on the world--sheet, $X^M( \tau,\sigma)$
are the string coordinates in AdS$_5$, and $g_{ab}$, with
$a,\,b=\tau,\,\sigma$, is the induced metric on the string world--sheet.

Choosing $\tau=t$ and $\sigma = z$ as the two coordinates parametrizing
the world-sheet we can write the embedding functions and its derivatives
as
 \beq
 X^M = (t,\brs,z),
 \qquad \dot{X}^M = (1,\dbrs,0),
 \qquad {X^M}' = (0,\pbrs,1),
 \eeq
where a dot or a prime on $\brs$ denote a derivative with respect to $t$
or $z$ respectively. The individual components and the determinant of the
induced metric read
 \beq\label{induced}
 g_{\tau\tau} &=& \dot{X} \cdot \dot{X} = |G_{00}|\,(-1 + \dbrs^2),
 \qquad % \nn*[0.15cm]
 g_{\sigma\sigma} = X' \cdot X' = |G_{00}|\,(1 + \pbrs^2),
 \nn*[0.15cm]
 g_{\tau\sigma} &=& \dot{X} \cdot X' = |G_{00}|\, \dbrs\cdot\pbrs,
 \qquad %\nn*[0.15cm]
 \hspace{0.37cm}\sqrt{-g} = |G_{00}|\, \sqrt{1-\dbrs^2+\pbrs^2 - (\dbrs
\times \pbrs)^2},
 \eeq
with $|G_{00}|=L^2/z^2$. The condition that the action \eqref{NG} be
stationary under small variations $\brs \to \brs + \delta\brs(t,z)$
yields the string equations of motion
 \beq\label{EOM}
 \frac{\del }{\del t}\,
 \frac{(1 + \pbrs^2)\dbrs -(\dbrs \cdot \pbrs) \pbrs}{\sqrt{1-\dbrs^2+\pbrs^2
- (\dbrs \times \pbrs)^2}}
 - \frac{1}{|G_{00}|}\,\frac{\del}{\del z} \, \frac{|G_{00}| [(1-
\dbrs^2)\pbrs +(\dbrs \cdot \pbrs) \dbrs]}
 {\sqrt{1-\dbrs^2+\pbrs^2 - (\dbrs \times \pbrs)^2}} = 0\,.
 \eeq
As shown by Mikhailov \cite{Mikhailov:2003er}, the general solution
$\brs(t,z)$ to the above equation is implicitly determined by
 \beq\label{EOMsolt}
 t \,=\, t_q + \gq z,\qquad
 \brs \,=\, \brq + \bvq(t-t_q)\,,
 \eeq
where $\brq$, $\bvq$ and $\gq$ are evaluated at $t_q$, with $\bvq\equiv
\dot{\bm r}_q$ the quark velocity and $\gq \equiv 1/({1 - \vq^2})^{1/2}$.
These two equations should be understood as follows: by solving the first
equation \eqref{EOMsolt}, one obtains $t_q$ as a function of $t$ and $z$,
which is then inserted into the second equation \eqref{EOMsolt} to obtain
the function $\brs(t,z)$. The solutions thus obtained must be restricted
to $z\ge z_m$. By assumption, $z_m$ is arbitrarily small, but the limit
$z_m\to 0$ can be taken only after performing the `ultraviolet
renormalization', {\em i.e.}, after absorbing a would--be divergent
contribution in that limit in the definition of the quark mass. To
clarify the physical interpretation of eqs.~\eqref{EOMsolt}, notice that
they imply
 \beq\label{rsrq}
 (\brs-\brq)^2+ z^2=(t-t_q)^2\qquad\mbox{and}\qquad
 t_q(t,z=0)=t\,,\qquad \brs(t,z=0)= \brq(t)\,,\eeq
and that the velocity of light in AdS$_5$ is equal to one within the
present conventions. Hence \eqn{rsrq} can be interpreted as follows: a
light signal emitted at time $t_q$ at the point with $\br=\brq$ and $z=0$
reaches the string (at the point with coordinates $\brs$ and $z$) at the
later time $t$. Thus, eqs.~\eqref{EOMsolt} show how the string gets built
within the bulk via radiation from the quark on the boundary.

We shall often need the derivatives of $t_q$ w.r.t. $t$ and $z$, which
read (below, $\baq\equiv\dot{\bm{v}}_q$)
 \beq\label{dtqdts}
 \frac{\del t_q}{\del t} = \frac{1}{1 + z \gq^3 \,\bvq \cdot \baq}\,,\quad
 \frac{\del t_q}{\del z} = -\frac{\gq}{1 + z \gq^3 \,\bvq \cdot \baq}\,.
 \eeq
Using these formul\ae{} together with eqs.~\eqref{EOMsolt} it is
straightforward to express the derivatives of $\brs$ in terms of the
boundary motion:
 \beq\label{dbrs of rq}
 \dbrs = \bvq + \frac{\gq z \baq}{1 + z \gq^3 \,\bvq \cdot \baq},
 \quad
 \pbrs = -\frac{\gq^2 z \baq}
 {1 + z \gq^3 \,\bvq \cdot \baq},
 \eeq
The following identity is also useful:
 \beq\label{ident}
 \sqrt{1-\dbrs^2+\pbrs^2 - (\dbrs \times \pbrs)^2} =
 \frac{1}{\gq}\,\frac{\del t_q}{\del t}\,.
 %,\qquad \pbrs + \gq \dbrs = \gamma_q \bvq.
 \eeq

\section{5D bulk stress tensor and backreaction}
\label{bulk}

The calculation of the space--time distribution of the energy produced by
the heavy quark requires to solve the `backreaction problem', that is, to
compute the perturbation $\delta G_{MN}$ of the AdS$_5$ metric associated
with the string. For the problem at hand, this perturbation is relatively
small, of $ \order{1/N_c^2}$, and can be computed by solving the
linearized Einstein equation with $t_{MN}$ (the string stress tensor) as
a source. The expectation value $\langle T_{\mu\nu}\rangle$ of the
physical stress tensor in the boundary gauge theory is then obtained from
the near--boundary ($z\to 0$) behaviour of $\delta G_{\mu\nu}$. Thus, it
becomes apparent that we first need to compute the 5D bulk tensor and
since our source is a string it will be proportional to $\delta^{(3)}(\br
- \brs)$. More precisely
 \beq\label{tMN}
 t^{MN}(t, \br, z)\, =\,
 -\frac{T_0}{\sqrt{-G}}\sqrt{-g}\,g^{ab}\,
 \partial_a X^M \,\partial_b X^N\,
 \delta^{(3)}(\br-\bm r_s)
 \equiv \tti_{MN}\, \delta^{(3)}(\br-\bm r_s),
 \eeq
where it is not hard to see that $\tti_{MN}$ is given by
 \beq
 \tilde{t}^{MN} = \frac{T_0}{\sqrt{-g} \sqrt{-G}}\,
 \left[
 g_{\sigma\sigma} \dot{X}^M \dot{X}^N +
 g_{\tau \tau} {X^M}' {X^N}' -
 g_{\tau \sigma} \left(\dot{X}^M {X^N}' + {X^M}' \dot{X}^N\right)
 \right].
 \eeq
Now we calculate the components of the above and at the same time lower
the indices. The metric is diagonal and we do this by multiplying each
component by $\pm |G_{00}|^2$, where we use the minus sign only when one
of the two indices is equal to 0. After substitution of the common
coefficient (notice that there is a third $|G_{00}|$ factor coming from
$g_ {ab}$, \eqnum{induced})
 \beq
 \frac{T_0 |G_{00}|^3}{\sqrt{-g}\sqrt{-G}} =
 \frac{\sqrt{\lambda}}{2 \pi}\, \frac{z}{L^3}\, \frac{1}
 {\sqrt{1-\dbrs^2+\pbrs^2 - (\dbrs \times \pbrs)^2}} =
 \frac{\sqrt{\lambda}}{2 \pi}\, \frac{z \gq}{L^3}\,\frac{\del t}{\del t_q},
 \eeq
we have
 \beq
 \tti_{00} &=& \frac{\sqrt{\lambda}}{2 \pi}\,
 \frac{z \gq}{L^3}\,\frac{\del t}{\del t_q}\,
 (1 + \pbrs^2),
 \\
 \tti_{0i} &=& \frac{\sqrt{\lambda}}{2 \pi}\,
 \frac{z \gq}{L^3}\,\frac{\del t}{\del t_q}\,
 [-(1 + \pbrs^2) \dot{x}_s^i + (\dbrs \cdot \pbrs) x_s^{\prime i}],
 \\
 \tti_{05} &=& \frac{\sqrt{\lambda}}{2 \pi}\,
 \frac{z \gq}{L^3}\,\frac{\del t}{\del t_q}\,
 \dbrs \cdot \pbrs,
 \\
 \tti_{ij} &=& \frac{\sqrt{\lambda}}{2 \pi}\,
 \frac{z \gq}{L^3}\,\frac{\del t}{\del t_q}\,
 [(1 + \pbrs^2)\dot{x}_s^i \dot{x}_s^j +
 (-1 + \dbrs^2) x_s^{\prime i} x_s^{\prime j}-
 (\dbrs \cdot \pbrs) (\dot{x}_s^i x_s^{\prime j} + \dot{x}_s^j x_s^{\prime
i})],
 \\
 \tti_{i5} &=& \frac{\sqrt{\lambda}}{2 \pi}\,
 \frac{z \gq}{L^3}\,\frac{\del t}{\del t_q}\,
 [(-1 + \dbrs^2) x_s^{\prime i} - (\dbrs \cdot \pbrs) \dot{x}_s^i],
 \\
 \tti_{55} &=& \frac{\sqrt{\lambda}}{2 \pi}\,
 \frac{z \gq}{L^3}\,\frac{\del t}{\del t_q}\,
 (-1 + \dbrs^2).
 \eeq
Expressing the components in terms of the boundary motion according to
\eqnum{dbrs of rq}, we finally deduce
 \beq\label{t00}
 \tti_{00} &=& \frac{\sqrt{\lambda}}{2 \pi}\,
 \frac{z \gq}{L^3}\,\frac{\del t_q}{\del t}
 \left\{ 1+ 2 z \gq^3\, \bvq \cdot \baq + z^2 \gq^6 [\baq^2 - (\bvq \times
\baq)^2] \right\},
 \\\label{t0i}
 \tti_{0i} &=& \frac{\sqrt{\lambda}}{2 \pi}\,
 \frac{z \gq}{L^3}\,\frac{\del t_q}{\del t}
 \left\{-z \gq a_q^i - (1+ 2 z \gq^3 \bvq \cdot \baq) \vq^i - z^2 \gq^6
[\baq^2 - (\bvq \times \baq)^2] \vq^i \right\},
 \\\label{t05}
 \tti_{05} &=& \frac{\sqrt{\lambda}}{2 \pi}\,
 \frac{z \gq}{L^3}\,\frac{\del t_q}{\del t}
 \left\{-  z \gq^2\, \bvq \cdot \baq - z^2 \gq^5 [\baq^2 - (\bvq \times \baq)
^2] \right\},
 \\\label{tij}
 \tti_{ij} &=& \frac{\sqrt{\lambda}}{2 \pi}\,
 \frac{z \gq}{L^3}\,\frac{\del t_q}{\del t}
 \big\{z \gq  (\vq^i a_q^j + \vq^j a_q^i)
 +(1 + 2 z \gq^3\, \bvq \cdot \baq) \vq^i\vq^j
 \nn
 && \phantom{\hspace*{2.2cm}}+ z^2 \gq^6 [\baq^2 - (\bvq \times \baq)^2]
\vq^i\vq^j \big\},
 \\\label{ti5}
 \tti_{i5} &=& \frac{\sqrt{\lambda}}{2 \pi}\,
 \frac{z \gq}{L^3}\,\frac{\del t_q}{\del t}
 \left\{z a_q^i + z \gq^2 \bvq \cdot \baq \vq^i + z^2 \gq^5 [\baq^2 - (\bvq
\times \baq)^2] \vq^i \right\},
 \\\label{t55}
 \tti_{55} &=& \frac{\sqrt{\lambda}}{2 \pi}\,
 \frac{z \gq}{L^3}\,\frac{\del t_q}{\del t}
 \left\{- \frac{1}{\gq^2} + z^2 \gq^4 [\baq^2 - (\bvq \times \baq)^2] \right
\}.
 \eeq

\section{Energy density in the gauge theory}
\label{back}

The details of this analysis relating the energy density $\mcal {E}
\equiv \langle T_{00}\rangle$ on the boundary to the string stress tensor
in the bulk can be found in Ref.~\cite{Athanasiou:2010pv} from which we
shall simply borrow the final formul\ae{}. Namely, one has
 \beq
 \mcal{E}(t,\br) \!&=&\! \EA(t,\br)+\EB(t,\br),
 \eeq
where the two contributions are
 \beq \label{EA}
 \EA \!&=&\! \frac{2 L^3}{\pi}\!
 \int\! \frac{\dif^4\rp \,\dif z}{z^2}\, \Theta(t-\tp) \delta''(\mcal{W})\!
 \left[z (2 t_{00} - t_{55}) - (t-\tp) t_{05} + (x - \xp)^i t_{i5}
 \right],
 \\ \label{EB}
 \EB\!&=&\! \frac{2 L^3}{3 \pi}\!
 \int\! \frac{\dif^4\rp \,\dif z}{z}\, \Theta(t-\tp) \delta'''(\mcal{W})\!
 \left[|\br - \brp|^2 (2 t_{00} -2 t_{55} + t_{ii}) -
 3 (x - \xp)^i (x - \xp)^j t_{ij} \right].
 \eeq
The argument of $t_{MN}$ is $(\tp,\brp,z)$ and the quantity
 \beq\label{W}
 \mcal{W} \equiv -(t-\tp)^2 + (\br - \brp)^2 +z^2\,
 \eeq
is proportional to the 5D invariant distance between the source point in
the bulk and the measurement point on the boundary.
Eqs.~\eqref{EA}--\eqref {EB} are essentially convolutions of $t_{MN}$
with the graviton bulk--to--boundary propagator.

The integration over $\dif^3\brp$ is trivially done using the
$\delta$-function of the string stress tensor given in \eqn{tMN} and then
by using eqs.~\eqref{t00}--\eqref{t55} we can express the integrands in
eqs.~\eqref{EA} and \eqref{EB} in terms of the boundary motion. Since
$\brq,\bvq$ and $\baq$ are evaluated at $t_q$, the calculation simplifies
if we change variable from $\tp$ to $t_q$. Then for the $\EA$ term we
have
 \beq\label{EAtemp}
 \EA = \frac{\sqrt{\lambda}}{\pi^2}
 \int \dif t_q\,\dif z\, \delta''(\mcal{W}_q + 2 \gq  \Xi z)
 [A_0(t_q) + z A_1(t_q) + z^2 A_2(t_q)],
 \eeq
with the definitions
 \beq\label{Wq}
 \mcal{W}_q \equiv -(t - t_q)^2 + |\br - \brq|^2,
 \qquad
 \Xi \equiv
 (t-t_q) - \bvq \cdot (\br - \brq)=
 \,\frac{1}{2}\,\frac{\dif \mcal{W}_q}{\dif t_q}\,,
 \eeq
and where the coefficients of the polynomial in $z$ in the square bracket
of \eqn{EAtemp} are
 \beq\label{A0}
 A_0 &=& 3 \gq -\gq\vq^2 + \gq^3 \bvq \cdot \baq [t- t_q +\bvq \cdot (\br -
\brq)] +
 \gq \,\baq \cdot (\br - \brq),
 \\*[0.15cm]\label{A1}
 A_1 &=& 2 \gq^4 \bvq \cdot \baq
 + \gq^6 [\baq^2 - (\bvq \times \baq)^2]
 [t- t_q +\bvq \cdot (\br - \brq)],
 \\*[0.15cm]\label{A2}
 A_2 &=& 0.
 \eeq
Note that, given the $z$--dependencies of the tensor components
$\tti_{MN}$ in eqs.~\eqref{t00}--\eqref{t55} and the other factors of $z$
in the integrand of \eqn{EA}, one would {\em a priori} expect the
integrand of \eqn{EAtemp} to contain a polynomial of second order in $z$.
However, in reality this polynomial is {\em linear} since, as shown in
\eqn{A2} above, the coefficient $A_2$ of the quadratic term is
identically zero, due to rather non--trivial cancelations. For example,
for the terms proportional to $z^2 [\baq^2 - (\bvq \times \baq)^2]$,
there are four different contributions yielding a total coefficient
$2\gq^2 - 1 -\gq^2 - \gq^2 \vq^2$, which indeed vanishes. This has
important consequences to which we shall shortly return.

After replacing $\tp\to t_q$ as the integration variable, the argument of
the $\delta$--function has become linear in $z$ and thus it is easier to
perform first the corresponding integration. The two derivatives in
$\delta''$ can be taken w.r.t. $\mcal{W}_q$ and pulled outside the
$z$--integration. Then the $\delta$--function sets
 \beq\label{zdelta}
 z = -\frac{\mW_q}{2 \gq \Xi}\,.
 \eeq
Since causality requires $\Xi>0$ and $z$ takes only non--negative values,
it is clear that the above result is non--zero only for $\mcal{W}_q\le
0$. Therefore,
 \beq
 \EA = \frac{\sqrt{\lambda}}{\pi^2}
 \int \dif t_q
 \left\{
 -\frac{A_1}{4 \gq^2 \Xi^2}\,
 \frac{\del^2}{\del \mW_q^2}\,
 [\Theta(-\mW_q) \mW_q]
 +\frac{A_0}{2 \gq \Xi}\,
 \frac{\del^2}{\del \mW_q^2}\,
 \Theta(-\mW_q)
 \right\}.
 \eeq
For the first term it is straightforward to compute the two derivatives.
In the second term, we first take one derivative, then rewrite the second
one as $\del/\del \mW_q  = (2 \Xi)^{-1} \del/\del_{t_q}$, and finally
integrate by parts to obtain
 \beq\label{inttq}
 \EA = \frac{\sqrt{\lambda}}{4 \pi^2}
 \int \dif t_q\,
 \delta(\mW_q)
 \left(
 \frac{A_1}{\gq^2 \Xi^2}+
 \frac{\del}{\del t_q}\,
 \frac{A_0}{\gq \Xi^2}
 \right).
 \eeq
Let $t_r=t_r(t, \br)$ denote the value of $t_q$ for which $\mW_q(t_q)=0$,
that is
 \beq\label{tr}
 t - t_r = |\br - \brq(t_r)|.
 \eeq
Writing $\delta(\mW_q) = \delta(t_q - t_r)/2 \Xi$ we finally arrive at
 \beq\label{EAfinal}
 \EA = \frac{\sqrt{\lambda}}{8 \pi^2}\,
 \frac{A_1}{\gq^2 \Xi^3}\,
 +\,
 \frac{\sqrt{\lambda}}{8 \pi^2}\,
 \frac{1}{\Xi}\,
 \frac{\del}{\del t_r}\,
 \frac{A_0}{\gq \Xi^2}\,.
 \eeq
One should be cautious to treat $t_r$ as a symbolic variable: only after
the derivative is performed, one can replace $t_r$ by its actual
dependence on $t$ and $\br$ via the solution to \eqn{tr}. Also, in the
previous manipulations we have been a little imprecise about the
integration limits in $t_q$ and $z$ after the change of variables and the
associated boundary terms. The most interesting case for us here will be
a situation where the motion keeps going for ever, meaning
$-\infty<t'<t$. Then \eqn{EOMsolt} implies $-\infty<t_q<t$ and
$0<z<\infty$, and one can easily check that the integration by parts
generates no boundary terms. Indeed, at the upper limit $t=t_q$ and the
constraint \eqn{tr} can be satisfied only when $\br = \brq(t)$, a
situation that we shall not consider. Also, at the lower limit $t_q\to
-\infty$, \eqn{tr} cannot be satisfied for any finite $\br$. Other
situations, where the integration domain for $t'$ is finite, need to be
considered case by case.

To summarize, the above integral over $t_q$ has support only at
$t_q=t_r$, where $\mW_q=0$, cf. \eqn{inttq}. Via \eqn{zdelta}, this
implies that the integration over $z$ receives contributions from the
endpoint at $z=0$ alone. These special properties are the consequence of
the above mentioned cancelation of the terms proportional to $z^2$ in the
integrand of \eqn{EAtemp}. In turn, they imply that the final result
\eqref{EAfinal} has {\em the same causal structure as the corresponding
classical result}, that is, as the energy density produced by a source
with trajectory $\brq(t_q)$ in a classical field theory. Indeed, the
condition $\mW_q=0$ is recognized as the classical retardation condition
for the propagation of a signal at the speed of light. In particular,
$t_r(t, \br)$ coincides with the classical `retarded time' --- the time
$t_q$ at which a light signal must be emitted by the source located at
$\brq$ in order to be received at the point $\br$ at some latter time
$t$. Furthermore, the space--time pattern of the energy in \eqn{EAfinal}
must be the same as in the corresponding classical problem, since this is
entirely fixed by the trajectory of the source together with the
condition that the signal propagates at the speed of light. In
particular, when focusing on the radiation part we expect no quantum
broadening to emerge.

For the calculation of $\EB$ we proceed similarly. Expressing the
integrand in terms of the boundary motion, and after some tedious but
straightforward algebra we arrive at
 \beq
 \hspace{-0.7cm}
 \EB = \frac{\sqrt{\lambda}}{\pi^2}
 \int \dif t_q\,\dif z\, \delta'''(\mcal{W}_q + 2 \gq  \Xi z)
 [B_0(t_q) + z B_1(t_q)+ z^2 B_2(t_q)+ z^3 B_3(t_q)+ z^4 B_4(t_q)],
 \eeq
where the coefficients are given by
 \beq\label{B0}
 \hspace{-0.4cm}
 B_0 &=& \frac{4}{3\gq}\,(\br - \brq)^2 + \gq [(\br - \brq) \times \bvq]^2,
 \\*[0.15cm]\label{B1}
 \hspace{-0.4cm}
 B_1 &=& -\frac{8}{3}\,\bvq \cdot (\br - \brq)
 + 2 \gq^2 [(\br - \brq) \times \bvq] \cdot
 \{(\br - \brq) \times\ [\gq^2 (\bvq \cdot \baq)\bvq + \baq ]\}\,,
 %=\frac{\del}{\del t_q}\Big(\gamma_q B_0\Big),
 \\*[0.15cm]\label{B2}
 \hspace{-0.4cm}
 B_2 &=& \frac{4}{3}\,\gq \vq^2
 - 2 \gq^3 [(\br - \brq) \times \bvq]\cdot(\bvq \times \baq) +
 \gq^7 [\baq^2 - (\bvq \times \baq)^2] [(\br - \brq) \times \bvq]^2,
 \\*[0.15cm]\label{B34}
 B_3 &=& B_4 =0.
 \eeq
{\em A priori}, the integrand can involve a quartic polynomial in $z$,
but in reality this polynomial is just quadratic, since the terms
proportional to $z^3$ and $z^4$ have exactly canceled among various
contributions. Thus the integration has similar properties to the one for
$\EA$, since we now have three derivatives to take. Once again, the
integrations over $z$ and $t_q$ are fixed by \eqn{zdelta} and
respectively the condition $\mW_q=0$, which together imply $t_q=t_r$ and
$z=0$. One finally obtains
 \beq\label{EBfinal}
 \EB = -\frac{\sqrt{\lambda}}{8 \pi^2}\,
 \frac{B_2}{\gq^3 \Xi^4}
 \,-\,
 \frac{\sqrt{\lambda}}{16 \pi^2}\,
 \frac{1}{\Xi}\,
 \frac{\del}{\del t_r}\,
 \frac{B_1}{\gq^2 \Xi^3}
 \,-\,
 \frac{\sqrt{\lambda}}{16 \pi^2}\,
 \frac{1}{\Xi}\,
 \frac{\del}{\del t_r}
 \left(
 \frac{1}{\Xi}\,
 \frac{\del}{\del t_r}\,
 \frac{B_0}{\gq \Xi^2}
 \right).
 \eeq
The same discussion as for \eqn{EAfinal} applies to potential boundary
terms. Eqs.~\eqref{EAfinal} and \eqref {EBfinal} are our final results
for the total energy density produced by the heavy quark. It is
interesting to notice that $B_0$ and $B_1$ are related as $B_1=\frac
{\del}{\del t_q}(\gamma_q B_0)$, so the last two terms in \eqnum{EBfinal}
can be combined to yield a somewhat simpler expression for $\EB$ :
 \beq\label{EBfinalv2}
 \EB = -\frac{\sqrt{\lambda}}{8 \pi^2}\,
 \frac{B_2}{\gq^3 \Xi^4}
 \,-\,
 \frac{\sqrt{\lambda}}{8 \pi^2}\,
 \frac{1}{\Xi}\,
 \frac{\del}{\del t_r}
 \left(
 \frac{1}{\gq \Xi^2}\,
 \frac{\del}{\del t_r}\,
 \frac{B_0}{\Xi}
 \right).
 \eeq

In the next section, we shall use these results to extract the energy
radiated by the heavy quark. But before that, let us perform a first,
non--trivial check of these formul\ae{} by using them to recover the
known result for the Coulomb energy of a heavy quark which moves at
constant velocity (and thus it does not radiate). Assuming uniform linear
motion with velocity $\upsilon$ along the $x$ axis, we have $\baq=0$ and
then the expressions for the coefficients $A_i$ and $B_i$ simplify
considerably. After simple manipulations, we deduce
 \beq\label{EAHQ1}
 \EA = \frac{\sqrt{\lambda}}{4 \pi^2}\,
 \frac{\gamma^2 (3-\upsilon^2) }{[\bxT^2 + \gamma^2 (x - \upsilon t)^2]^2}
 \,.
 \eeq
and respectively
  \beq\label{EBHQ}
 \EB = - \frac{\sqrt{\lambda}\gamma^2}{6 \pi^2}\,
 \frac{(4- 2\upsilon^2)[\bxT^2 + \gamma^2 (x - \upsilon t)^2] +
 \upsilon^2\gamma^2 (x - \upsilon t)^2}{[\bxT^2 + \gamma^2
 (x - \upsilon t)^2]^3}.
 \eeq
which add together to the expected result
\cite{Gubser:2007xz,Chesler:2007an,Dominguez:2008vd} :
 \beq\label{uniform}
 \mcal{E} = \frac{\sqrt{\lambda}\gamma^2}{12 \pi^2}\,
 \frac{(1+ \upsilon^2)\bxT^2 + (x - \upsilon t)^2}
 {[\bxT^2 + \gamma^2 (x - \upsilon t)^2]^3}.
 \eeq
Note that all the three terms in \eqn{EBfinal} for $\EB$ contribute to
\eqn{EBHQ}, while \eqn{EAHQ1} receives contributions only from the
second, derivative, term in \eqn{EAfinal}.

\section{Radiated energy and power}
\label{rad}

From now on we shall focus on the part of the energy density which is
radiated. Following the standard definition in the literature, we shall
identify the radiation as the part of the energy density which falls like
$1/R^2$ (with $\bm{R} \equiv \br - \brq(t_r)$) at large distances from
the source. For this definition to be meaningful, we shall consider only
observation points $\br$ which are sufficiently far away from the
position $\brq(t)$ of the quark at the observation time $t$ for the
dominant contribution of the energy density at $\br$ to be falling like
$1/R^2$. (Indeed, if $\br$ is relatively close to $\brq(t)$, then the
retardation condition \eqref{tr} allows for solutions $t_r$ with
$t_r\simeq t$ and $\brq(t_r)\simeq \br$, and then the energy density at
$\br$ is dominated by the near--field of the heavy quark, {\em i.e.} by
its Coulomb energy, and not by radiation.)

To make the power counting with respect to $1/R$ more transparent, it is
useful to notice that, for $t_q=t_r$ obeying \eqn{tr}, one has $\Xi = R(1
- \bm{n} \cdot \bvq)$, where we have defined $\bm{n}$ as the unit vector
along $\bm{R}$. Then, by inspection of the expressions in the previous
section, one can check that, first, the pieces showing the slowest decay
at large distances in Eqs.~\eqref{EAfinal} and \eqref {EBfinal} are those
which behave like $1/R^2$, as expected, and, second, in order to isolate
these pieces, it is enough to keep the terms in the coefficients $A_i$
which are proportional to $R$ or $t-t_q$ and the terms in the
coefficients $B_i$ which are proportional to $R^2$. By doing that, one
eventually finds that radiative contributions $\propto 1/R^2$ to the
energy density come from all the terms in Eqs.~\eqref{EAfinal} and \eqref
{EBfinal} which are proportional to either the square of the
acceleration, or to its time derivative (also known as the `jerk'). Thus,
the radiation vanishes in the absence of acceleration, as expected.
However, unlike what would happen in a classical theory, or in a weakly
coupled theory at leading order, where the radiation involves {\em only}
terms proportional to the square of the acceleration (see e.g. the
discussion in Sec.~\ref{classical} below), in the present calculation at
strong coupling we also find contributions proportional to the jerk
$\dot{\bm a}_q$.

In what follows, we shall exhibit all the radiative contributions to the
energy density in Eqs.~\eqref{EAfinal} and \eqref {EBfinal}. In turns out
that, in view of the subsequent physical discussion and also of the
comparison with the respective classical results in Sec.~\ref{classical},
it is meaningful to separate between two types of such contributions:
\texttt{(i)} those generated by terms in the string stress tensor which
are by themselves proportional to the square of the acceleration, and
\texttt{(ii)} those coming from the terms in Eqs.~\eqref{EAfinal} and
\eqref {EBfinal} which involve derivatives w.r.t. $t_r$.

\texttt{(i)} Contributions proportional to the acceleration squared, more
precisely to the structure $\baq^2 - (\bvq \times \baq)^2$, and which are
originating from the components \eqref{t00}--\eqref{t55} of $\tti_{MN}$,
are visible in \eqn{A1} for $A_1$ and in \eqn{B2} for $B_2$. They
contribute to the energy density via the terms without derivatives in
Eqs.~\eqref{EAfinal} and \eqref {EBfinal}, and yield
 \beq\label{EAB1rad}
 \EA^{\rm (1)} &=&
 \frac{\sqrt{\lambda}}{8 \pi^2}\,
 \frac{\gq^4 [\baq^2 - (\bvq \times \baq)^2]}{R^2}\,
 \frac{1 + \bm{n} \cdot \bvq}{(1 - \bm{n} \cdot \bvq)^3}\,,
 \\
 \EB^{\rm (1)} &=&
 -\frac{\sqrt{\lambda}}{8 \pi^2}\,
 \frac{\gq^4 [\baq^2 - (\bvq \times \baq)^2]}{R^2}\,
 \frac{(\bm{n} \times \bvq)^2}{(1 - \bm{n} \cdot \bvq)^4}\,.
 \eeq
These two contributions combine to give the following, relatively simple,
expression
 \beq\label{Erad1}
 \mcal{E}_{\rm rad}^{(1)}(t,\br)\, =\,
 \frac{\sqrt{\lambda}}{8 \pi^2}\,
 \frac{\gq^2 [\baq^2 - (\bvq \times \baq)^2]}{(\br - \brq)^2 (1 - \bm{n}
\cdot \bvq)^4}\,,
 \eeq
where it is understood that all quantities related to the motion of the
quark ($\brq$, $\bvq$, and $\baq$) are evaluated at $t_q=t_r(t,\br)$.
Note that the other terms in $A_1$ and $B_2$ do not generate
contributions of order $1/R^2$ to the energy density.

\texttt{(ii)} The remaining terms of order $1/R^2$ arise from the 3rd and
4th term of $A_0$, from $B_0$, and from the 2nd term of $B_1$. {\em A
priori}, that is, within the coefficients $A_i$ and $B_i$, these terms
depend only upon the quark velocity $\bvq$ and are at most linear in the
acceleration $\baq$, but after taking the derivatives w.r.t. $t_r$ in
eqs.~\eqref{EAfinal} and \eqref{EBfinal}, they generate contributions
proportional to the square of the acceleration, or to its derivative.
Defining $\xi = 1 - \bm{n} \cdot \bvq$, we find
 \beq\label{EA2}
 \EA^{(2)} &\!=\!& \frac{\sqrt{\lambda}}{8 \pi^2 R^2 \xi}\,
 \frac{\del}{\del t_r}
 \left[\frac{\bm{n}\cdot \baq}{\xi^2}
 + \frac{\gq^2 \bvq \cdot \baq (2 - \xi)}{\xi^2}\right]\,
 \\\label{EB2}
 \EB^{(2)} &\!=\!& -\frac{\sqrt{\lambda}}{8 \pi^2 R^2 \xi}\,
 \frac{\del}{\del t_r}
 \left[-\frac{\bm{n}\cdot \baq (1 - \xi)}{\xi^3}
 + \frac{\gq^2 \bvq \cdot \baq (2 - \xi)}{\xi^2} +
  \frac{1}{\xi}\,\frac{\del}{\del t_r}
  \left(\frac{1}{6 \gq^2 \xi^2} + \frac{1}{\xi}\right) \right],
 \eeq
where we have neglected derivatives acting on $\bm{R}$ or $t-t_q$ since
they generate terms  which fall faster than $1/R^2$. Performing the
derivative on $1/\xi$ of the  last term in Eq.~\eqref{EB2} (we do not
differentiate the unit vector $\bm{n} $ since this would lead again to
terms falling faster than $1/R^2$~; that is, we use $\del_{t_r} \xi
\simeq - \bm{n} \cdot \baq$), we see that $\EB^{(2)}$ becomes
 \beq\label{EB22}
 \EB^{(2)} = -\frac{\sqrt{\lambda}}{8 \pi^2 R^2 \xi}\,
 \frac{\del}{\del t_r}
 \left[\frac{\bm{n}\cdot \baq}{\xi^2}
 + \frac{\gq^2 \bvq \cdot \baq (2 - \xi)}{\xi^2} +
 \frac{1}{6\xi}\,\frac{\del}{\del t_r}
  \frac{1}{\gq^2 \xi^2}\right].
 \eeq
Thus, adding the two contributions from eqs.~\eqref{EA2} and \eqref{EB22}
we are left only with the last term in the last equation, which, after
preforming the first derivative w.r.t. $t_r$ and returning to the
original variables, is finally rewritten as
 \beq\label{Erad2}
 \mcal{E}_{\rm rad}^{(2)}(t,\br)\,=\,
 \frac{\sqrt{\lambda}}{24 \pi^2}\,
 \frac{1}{|\br - \brq|^2 (1 - \bm{n} \cdot \bvq)}\,
 \frac{\del}{\del t_r}
 \left[
 \frac{\bvq \cdot \baq}{(1 - \bm{n} \cdot \bvq)^3}
 - \frac{\bm{n} \cdot \baq}{\gq^2(1 - \bm{n} \cdot \bvq)^4}
 \right].
 \eeq
If one also performs the remaining derivative w.r.t. $t_r$, one finds
that all the ensuing terms are proportional to either the square or the
derivative of the acceleration, as anticipated.

The radiated energy density $\mcal{E}_{\rm rad}$ given by the sum of
eqs.~\eqref{Erad1} and \eqref {Erad2} is the main result of this paper.
The (relatively high) powers of $1 - \bm{n} \cdot \bvq$ visible in the
denominators of these expressions have a kinematical origin: they express
the angular collimation of the radiation due to the Lorentz boost, which
was to be expected, independently of the value of the coupling. In the
ultrarelativistic limit $v_q\simeq 1$ or $\gamma_q\gg 1$, one can write
(with $\alpha$ denoting the angle between the vectors $\bvq$ and
$\bm{n}$): $1 - \upsilon_q \cos\alpha\simeq (1/2)(1/\gamma^2_q +
\alpha^2)$. This makes it clear that the radiation is emitted within a
small angle $\alpha\sim 1/\gamma_q$ around the direction of the quark
velocity $\bvq(t_r)$, so like for the corresponding classical problem
\cite{Jackson}.

Using the above results for $\mcal{E}_{\rm rad}=\mcal{E}_{\rm
rad}^{(1)}+\mcal{E}_{\rm rad}^{(2)}$, we shall now compute the radiated
power. By integrating the energy conservation law $\del_t\langle
T^{00}\rangle +\del_i \langle T^{0i}\rangle=0$ over the whole space and
using Gauss' theorem together with $\langle T^{0i}\rangle \approx n^i
\langle T^{00}\rangle$ for the dominant respective contributions,
proportional to $1/R^2$, at large distances, one finds (recall the
notation $\langle T^{00}\rangle=\mcal{E}$)
 \beq
 -\,\frac{\dif E}{\dif t}\,=\,\lim_{r\to\infty} \,r^2\int\dif \Omega\
 \mcal{E}(t,\br)\,,\eeq
where the left hand side represents the energy radiated per unit of {\em
observation} time. In practice, it is more convenient to define the power
$P_{\rm rad}$ as the energy radiated per unit of {\em emission} time
$t_r$. Then by using the above formula together with ${\dif t}/{\dif
t_r}=\xi$, one sees that the power radiated per unit solid angle reads
(below, $R\to\infty$)
 \beq\label{P}
 \frac{\dif P_{\rm rad}}{\dif \Omega}= \frac{\dif t}{\dif t_r}\,
 R^2
 \mcal{E}_{\rm rad}
 \,\,\,\Rightarrow\,\,\,
 P_{\rm rad} =
 \int \dif \Omega \,(1 - \bm{n} \cdot \bvq) \,R^2 \mcal{E}_{\rm
rad}.
 \eeq
Note an important, implicit, assumption in the above argument: we have
made the hypothesis that, at all the points on a sphere at infinity
($r\to\infty$), we have $R\equiv |\br - \brq(t_r)|\simeq r$ (which in
turn implies that only the far--zone contributions $\propto 1/R^2$ to the
energy density and flux have to be retained). This is correct provided
the motion of the quark is {\em bounded}, such that its trajectory
$\brq(t_r)$ does not cross the sphere at infinity. More precisely, it is
enough that this condition be satisfied during the {\em acceleration
phase} of its motion, since this is the only phase which creates
radiation. Thus, our subsequent results for $P_{\rm rad}$ are only valid
provided there exists some fixed, but arbitrary, distance $r_0$ such that
$\brq(t_q)\le r_0$ for any $t_q$ within the acceleration phase. We shall
later make some comments on the case of unbounded motion.

Some standard and useful integrals to perform the angular integrations in
\eqn{P} are listed in Appendix \ref{integrals}. For the first
contribution coming from $\mcal{E}_{\rm rad}^{(1)}$ we find
 \beq\label{Prad1}
 P_{\rm rad}^{(1)}
 %=\frac{\sqrt{\lambda}}{8\pi^2}\,
 %\gq^2\,[\baq^2 - (\bvq \times \baq)^2]
 %\underbrace{\int \frac{\dif \Omega}{(1 - \bm{n} \cdot \bvq)^3}}_
 %{\displaystyle{4 \pi \gq^4}}
 =\frac{\sqrt{\lambda}}{2 \pi}
 \,\gq^6\,[\baq^2 - (\bvq \times \baq)^2]\,,
 \eeq
which is the result inferred in \cite{Mikhailov:2003er} from a
world--sheet analysis ({\em i.e.}, without an explicit calculation of the
backreaction, but merely via a calculation of the energy flux down the
string). Remarkably, this expression has the same structure as the
respective classical result, that is, the Li\'enard formula in classical
electrodynamics \cite{Jackson}, that we shall extend to the case of the
${\mathcal N}=4$ SYM theory in Sec.~\ref{classical}. For the second
contribution from $\mcal{E}_{\rm rad}^{(2)}$ we obtain
 \beq\label{Prad2}
  P_{\rm rad}^{(2)}
 %=\frac{\sqrt{\lambda}}{24\pi^2}\,
 %\frac{\del}{\del t_r}\left[
 %\left(\bvq \cdot \baq - \frac{1}{3\gq^2}\,\frac{\del}{\del t_r} \right)
 %\int \frac{\dif \Omega}{(1 - \bm{n} \cdot \bvq)^3}\right]
 = - \frac{\sqrt{\lambda}}{18 \pi}\,
 \frac{\del}{\del t_r}\, \gq^4 \bvq \cdot
 \baq.
 \eeq
The fact that this term is a total derivative w.r.t. the emission time
$t_r$ rises some puzzles for its interpretation as a contribution to the
radiated energy (see the discussion at the end of
Sect.~\ref{applications}). It is therefore interesting to notice that a
term with a similar structure has been interpreted in
Ref.~\cite{Chernicoff:2009re,Chernicoff:2009ff} as a contribution to the
{\em proper} energy of the quark (and not to its radiation). We shall
return to this point in the next section.

Eqs.~\eqref{Prad1} and \eqref{Prad2} are our final results for the
radiated power. It is important to keep in mind that these results have
been derived here for the case of a bounded quark motion. Because of
that, in evaluating these formul\ae{} one can use the approximation
$t_r\simeq t-r$ for most (but not all) purposes.

\section{Applications}
\label{applications}

Here we shall apply the general results derived in the previous section
for the radiated energy density and the power to specific quark motions.

\noindent {\bf (i) Uniform rotation:} We shall first recover the results
for uniform circular motion originally obtained in
\cite{Athanasiou:2010pv}. Using spherical coordinates $\br =
(r,\theta,\phi)$ and parametrizing the boundary motion as
 \beq
 \brq(t_r)=(R_0, \pi/2,\omega t_r),
 \eeq
our expressions in \eqn{Erad1} and \eqn{Erad2} lead to the following two
contributions to the density of the radiated energy
 \beq\label{Erot1}
 \mcal{E}_{\rm rad}^{(1)} &=& \frac{\sqrt{\lambda}}{8 \pi^2}\,
 \frac{a^2}{r^2 \xi^4}\,,
 \\ \label{Erot2}
 \mcal{E}_{\rm rad}^{(2)} &=& \frac{\sqrt{\lambda}\,
 \omega^2}{24 \pi^2 r^2}\,
 \frac{4 - 7 \xi -4 \upsilon^2 \sin^2\theta + 3 \xi^2}{\gamma^2 \xi^6},
 \eeq
with $\upsilon = \omega R_0$, $a=\omega^2 R_0$, and where, according to
our earlier definition above \eqn{EA2}, we have
 \beq
 \xi = 1 - \upsilon \sin\theta\sin(\phi - \omega t_r).
 \eeq
Note that the contribution in \eqnum{Erot2} is fully arising from the
last term, proportional to $\bm{n} \cdot \baq$, in \eqn{Erad2}, since
$\bvq \cdot \baq=0$ for the problem at hand.

Adding the two pieces in eqs.~\eqref{Erot1} and \eqref{Erot2} we find
 \beq\label{Erotsum}
 \mcal{E}_{\rm rad}
 = \frac{\sqrt{\lambda}\omega^2}{24 \pi^2 r^2}\,
 \frac{4 - 7 \xi -4 \upsilon^2 \sin^2\theta + 3 \gamma^2 \xi^2}
 {\gamma^2\xi^6},
 \eeq
which is indeed the same as the result for the radiated energy density in
\cite{Athanasiou:2010pv} (cf.~eq.~(3.72) there). The energy density
\eqref{Erotsum} is proportional to $1/\xi^6$, hence in the
ultrarelativistic limit $v\simeq 1$ or $\gamma\gg 1$, it is strongly
peaked at the minima of $\xi$, defined by $\sin\theta\sin(\phi - \omega
t_r)=1$. This condition describes a spiral in the plane $\theta=\pi/2$,
located at (recall that $t_r\simeq t-r$)
 \beq
 \phi(t,r)\,\simeq\,\frac{\pi}{2} +\omega(t-r)\,.
 \eeq
Using $1 - \upsilon \sin\theta\simeq (1/2)(1/\gamma^2 + \alpha^2)$ where
$\alpha\equiv {\pi}/{2}-\theta$, it is obvious that the energy is
localized within a small angle $\alpha\sim 1/\gamma$ around $\alpha=0$,
or $\theta=\pi/2$. This is the expected collimation due to the Lorentz
boost, as already discussed in relation with the general formul\ae{}
\eqref{Erad1} and \eqref {Erad2}. Furthermore, using \eqref{Erotsum} one
can check that the arms of the spiral have a tiny radial
width\footnote{In order to deduce this property from \eqn{Erotsum}, it is
not enough to use the simplified version of the retardation time
$t_r\simeq t-r$; rather, one needs a more precise analysis of the
retardation condition, which also takes into account the collimation of
the radiation along the direction of emission; see
\cite{Athanasiou:2010pv,Jackson}.} $\Delta r\sim R_0/\gamma^3$
\cite{Athanasiou:2010pv}, exactly like in the corresponding classical
problem \cite{Jackson}. As explained in the Introduction, this feature is
very surprising in a quantum theory at strong coupling, where one would
rather expect broadening due to the virtual quantum fluctuations. In the
context of our calculation, this follows from the fact that, as explained
on Sec.~\ref{back}, the whole backreaction arises from the string
endpoint at $z=0$. (See also \cite{Hubeny:2010bq} for a different
perspective of this problem.)

We now compute the radiated power in terms of the quark's own time. From
\eqn{Prad2}, it is clear that $P_{\rm rad}^{(2)}=0$ in this case, so the
only contribution comes from \eqn{Prad1} and reads
 \beq\label{Prot}
 P_{\rm rad} =
 \frac{\sqrt{\lambda}}{2 \pi}\, \gamma^4 a^2\,.
 \eeq
This coincides with the respective result in \cite{Athanasiou:2010pv} and
also with the result of the world--sheet analysis in
\cite{Mikhailov:2003er}.

\bigskip
\noindent {\bf (ii) Non--uniform circular motion:} It is of course
straightforward to apply our general expressions to an arbitrary circular
motion, but to be more precise we shall focus on the specific motion
 \beq\label{accrot}
 \brq(t_r) = (R_0,\pi/2,\phi_q(t_r)) \quad \mathrm{with} \quad \phi_q(t_r) =
 \sqrt{t_r^2 + b^2}/R_0\,,
 \eeq
for which we shall directly compute the radiated power. The (angular)
velocity, whose magnitude approaches the speed of light at large times,
and the acceleration are given by (below $\hat{e}_{r}$ and
$\hat{e}_{\phi}$ are the respective unit vectors)
 \beq\label{angva}
 \bvq(t_r) = \frac{t_r}{\sqrt{t_r^2 + b^2}}\, \hat{e}_{\phi}
 \qquad \mathrm{and} \qquad
 \baq = -\frac{1}{R_0}\,\frac{t_r^2}{t_r^2 + b^2}\, \hat{e}_{r}
 +\,\frac{b^2}{(t_r^2 + b^2)^{3/2}}\, \hat{e}_{\phi},
 \eeq
so in particular $\gamma_q=\sqrt{t_r^2 + b^2}/b$ and $\bvq \cdot \baq$ is
not vanishing anymore (in contrast to the case of uniform rotation).
Hence both terms contributing to the radiated power, \eqref{Prad1} and
\eqref{Prad2}, are now non--zero, and this is interesting as it allows us
to observe an hierarchy among these terms, that we believe to be generic.
Namely, $P_{\rm rad}^{(1)}$ dominates over $P_{\rm rad}^{(2)}$ in the
ultrarelativistic limit $\gq\gg 1$, and hence also for sufficiently large
times (in the problems where the velocity grows with time, due to
acceleration). Specifically, for the motion in \eqn{accrot}, one finds
 \beq\label{Pradaccrot}
 P_{\rm rad}^{(1)} = \frac{\sqrt{\lambda}}{2\pi}\,
 \frac{t_r^4 + b^2 R_0^2}{b^4
 R_0^2}
 \qquad \mathrm{and} \qquad
 P_{\rm rad}^{(2)} = -\frac{\sqrt{\lambda}}{18 \pi}\,\frac{1}{b^2}.
 \eeq
As anticipated, the first term $P_{\rm rad}^{(1)}$ is the dominant one
for large enough times such as $t_r^2\gg bR_0$. This example also gives
us some insight into the physical origin of this hierarchy: \eqn{Prad1}
for $P_{\rm rad}^{(1)}$ involves the component of the acceleration which
is {\em transverse} to the velocity\footnote{More precisely, the term
$\baq^2$ in \eqn{Prad1} receives contributions from both the radial and
the azimuthal components of the acceleration in \eqn{angva}, but the
dominant contribution at large times, represented by the term
proportional to $t_r^4$ in the numerator of $P_{\rm rad}^{(1)}$ in
\eqnum{Pradaccrot}, is generated by the radial piece of $\baq$.} (the
radial component of $\baq$ in \eqn{angva}), whereas $P_{\rm rad}^{(2)}$
in \eqn{Prad2} rather involves the respective {\em longitudinal}
component (tangential in the case of \eqn{angva}). We thus recover a
feature familiar in the context of classical radiation \cite{Jackson}:
rotation is much more effective than tangential acceleration in producing
radiation, since the velocity $\bvq$ changes rapidly in direction while
the particle rotates, even though its change in magnitude is relatively
small (or even zero for uniform rotation).

\bigskip

\noindent {\bf (iii) Uniform linear acceleration:} A classical particle
subjected to a constant force ${\bm F}=F\hat{e}_1$ follows a trajectory
$x_q(t_r)  = \sqrt{t^2_r + b^2}$ where $x\equiv x^1$, $b=m/F$ and we
selected convenient initial conditions at $t=0$. Clearly, this motion is
unbounded: the trajectory will eventually cross the sphere `at infinity'
that we use to define the radiated power. In view of that, we do not
expect our previous results for the power to also cover this case.
Notwithstanding, let us first see what these results yield if naively
applied to this case. Using
 \beq\label{uniform acc}
 v_q = \frac{t_r}{\sqrt{t^2_r + b^2}}, \qquad
 \gq = \frac{\sqrt{t^2_r + b^2}}{b}, \qquad a_q = \frac{b^2}
 {(t^2_r + b^2)^{3/2}},
 \eeq
and hence $\gq^4 \upsilon_q a_q = t_r/b^2$, one easily finds that the two
contributions in \eqn{Prad1} and respectively \eqn{Prad2} are now of the
same parametric order and thus contribute on the same footing to the
final result for the power, in contradiction with our general
expectations and also with the previous examples. Namely, one (naively)
has
 \beq
 P_{\rm rad}^{(1)} = \frac{\sqrt{\lambda}}{2 \pi b^2}
 \qquad \mathrm{and} \qquad
 P_{\rm rad}^{(2)} = -\frac{\sqrt{\lambda}}{18 \pi b^2}
 \quad\Longrightarrow\quad
 P_{\rm rad} = \frac{4\sqrt{\lambda}}{9 \pi b^2}\,.
 \eeq
Moreover, this result also contradicts the independent calculation in
Refs.~\cite{Dominguez:2008vd,Xiao:2008nr}, where the radiated power has
been extracted from a world--sheet analysis, as the energy flow across
the induced horizon at $z=b$. That previous calculation furnished a
result equal to $P_{\rm rad}^{(1)}$ in the above equation, which if
course would be also the prediction of Mikhailov's analysis for the
problem at hand \cite{Mikhailov:2003er} (since in that analysis the total
power reduces to our $P_{\rm rad}^{(1)}$). Clearly, these mismatches shed
further doubts on the validity of the above calculation, that is anyway
transgressing the validity limits of our general calculation.

Let us therefore redo our analysis of this problem, but in such a way to
stay within the limits of our general discussion. To that aim, we assume
that the quark is under uniform acceleration only for a finite period of
time $t_0$, and we measure the radiated energy at a distance $r \gg t_0$.
By doing this we effectively reduce the motion to a bounded one. Let us
be more specific and consider the one-dimensional motion
 \beq\label{lint0}
 x_q(t_r) = \Theta(-t_r)\, b +
 \Theta(t_r) \Theta(t_0-t_r) \sqrt{t_r^2 + b^2}
 + \Theta(t_r-t_0)\,\frac{t_0 t_r + b^2}{\sqrt{t_0^2 + b^2}},
 \eeq
where $t_0$ can be taken to be much larger than $b$ so that the quark
becomes eventually ultra-relativistic. (The last term in \eqn{lint0}
describes a constant velocity motion with the velocity acquired at
$t_r=t_0$.) In this last example we shall evaluate the total energy
radiated, and therefore we need to integrated the total power over $t_r$.
Then it is straightforward to see that \eqn{Prad2} will not contribute to
the final results, since it involves a total derivative w.r.t. $t_r$ and
the acceleration vanishes outside the interval $[0,t_0]$. Thus the power
in \eqn{Prad1} will determine the total energy radiated which is
 \beq
 E_{\rm rad} = \frac{\sqrt{\lambda}}{2\pi}\,\frac{t_0}{b^2},
 \eeq
in agreement with \cite{Dominguez:2008vd,Xiao:2008nr}.

This last calculation also illustrates a rather curious feature of
$P_{\rm rad}^{(2)}$ in \eqn{Prad2}, which makes us feel uncomfortable
about its physical interpretation as radiation: this term yields no
contribution to the radiated energy for any motion where the acceleration
is non--zero over only a finite interval of time, and also for any
periodic motion for which one computes the total radiation over one
period, or an integer multiple of it. As already mentioned after
\eqn{Prad2}, a term with a similar structure appears in the world--sheet
calculation of the total energy of the moving quark (which is the same as
the total energy carried by the string) in
Ref.~\cite{Chernicoff:2009re,Chernicoff:2009ff}. Specifically, by taking
the limit of a very heavy quark ($z_m\to 0$) in eqs.~(2.32)--(2.33) of
Ref.~\cite{Chernicoff:2009ff}, one obtains the following expression for
the quark energy
 \beq\label{Eqtot}
 E_q(t)\,=\,m_q\gamma_q(t)\,-\,\frac{\sqrt{\lambda}}{2\pi}\,
\gq^4 \bvq \cdot \baq\,+\,\frac{\sqrt{\lambda}}{2\pi}\int_{-\infty}^t\dif
 t_q\,\gq^6\,[\baq^2 - (\bvq \times \baq)^2]\,, \eeq
where the first two terms in the r.h.s. are interpreted
\cite{Chernicoff:2009re,Chernicoff:2009ff} as the quark proper (or
kinetic) energy, while the third one, which is the time--integral of
$P_{\rm rad}^{(1)}$ in \eqn{Prad1}, as the radiation. As anticipated, the
second term in the above equation has the same structure as the
contribution to the `radiated energy' that would be obtained from $P_{\rm
rad}^{(2)}$, \eqn{Prad2}, but with a different numerical coefficient.
Recall that the reason why we have identified this term as radiation in
Sect.~\ref{rad} was because it arises from a piece in the energy density
which falls off like $1/R^2$ at large $R$. This suggests the interesting
possibility that a piece of the quark proper energy have a tail at large
$R$ which cannot be distinguished from radiation. We leave this question,
as well the calculation of the total energy via the backreaction, for
further studies.

\section{The classical result}
\label{classical}

%We have thus seen that, in both calculations, the radiation propagates at
%the speed of light, so the energy density has the space--time
%localization, as determined by classical causality. A

In the previous discussion, we have already anticipated some similarities
between the predictions of the supergravity approximation for the strong
coupling limit and the corresponding results in the classical
approximation, which are also the leading order results at weak coupling.
For this comparison to be more precise and in preparation of the physical
discussion in the next section, in this section we shall explicitly solve
the corresponding classical problem --- the radiation by a heavy quark
undergoing some arbitrary motion in the ${\mathcal N}=4$ SYM theory at
weak coupling. To our knowledge, the general result that we shall derive
here has not been presented elsewhere, except for the case of the uniform
circular motion that was discussed in \cite{Athanasiou:2010pv}. But even
in that case, the final results and the associated physical discussion
have been plagued by some mistakes in the numerical factors that we shall
here correct.

The general structure of the classical theory describing a massive test
quark\footnote{More precisely an infinitely massive spin--1/2 particle
from the ${\mathcal N}=2 $ hypermultiplet, that is in the fundamental
representation of the SU$(N_c)$ gauge group.} propagating through the
vacuum of the ${\mathcal N}=4$ SYM theory has been clarified in Refs.
\cite{Chesler:2009yg,Athanasiou:2010pv}. As explained there, the heavy
quark radiates both vector (gauge) fields and scalar fields, and in the
limits of interest here (arbitrarily weak coupling and very large quark
mass) this radiation is described by decoupled, linear equations, which
generalize Maxwell equations to the theory at hand. These equations are
then solved in the standard way, to give
 \beq
 A^{\mu} = \frac{e_{\rm eff}}{4 \pi (1 - \bm{n} \cdot \bvq) R}\,(1,\bvq)\,
 \quad \mathrm{and} \quad
 \chi = \frac{e_{\rm eff}}{4 \pi \gamma_q (1 - \bm{n} \cdot \bvq) R}\,,
 \eeq
with $A^{\mu}$ the vector field, $\chi$ the scalar field, $\bm{R} = \br -
\brq$ and $\bm{n}$ the unit vector along $\bm{R}$ as earlier, and the
proper counting of the color degrees of freedom for the radiated field is
encoded in $e_{\rm eff}^2 \equiv \lambda/2$. As before, the above
expressions are to be evaluated at the retarded time $t_r$ which is the
solution to \eqn{tr}. In general the energy density is obtained from
 \beq
 \mcal{E}_{\rm vector} = \frac{1}{2}\left(\bm{E}^2 + \bm{B}^2\right)
 \quad \mathrm{and} \quad
 \mcal{E}_{\rm scalar} = \frac{1}{2}\left[(\del_t \chi)^2 +
 (\nabla \chi)^2\right],
 \eeq
with $\bm{B}$ the magnetic field. Since we are interested in the radiated
energy, we keep only the contributions which fall like $1/R$ when
computing the electric field and the derivative with respect to time of
the scalar field. This yields
 \beq\label{Efield}
 \bm{E}_{\rm rad} = \frac{e_{\rm eff}}{4 \pi R}\,
 \left[- \frac{\baq}
 {(1 - \bm{n} \cdot \bvq)^2}
 +\frac{(\bm{n} \cdot \baq)(\bm{n} - \bvq)}
 {(1 - \bm{n} \cdot \bvq)^3}\right],
 \\ \label{dtchi}
 (\del_t \chi)_{\rm rad} = \frac{e_{\rm eff}}{4 \pi R}\,
 \left[- \frac{\gamma_q \bvq \cdot \baq}
 {(1 - \bm{n} \cdot \bvq)^2}
 +\frac{\bm{n} \cdot \baq}
 {\gamma_q(1 - \bm{n} \cdot \bvq)^3}\right ].
 \eeq
Since moreover $|\bm{B}_{\rm rad}| = |\bm{E}_{\rm rad}|$ and $|(\del_t
\chi)_{\rm rad}| = |(\nabla \chi)_{\rm rad}|$ for the radiation, we
deduce that
 \beq\label{Evec}
 \mcal{E}_{\rm vector} &=&
 \frac{\lambda}{32 \pi^2 R^2}
 \left[\frac{\baq^2}{(1 - \bm{n} \cdot \bvq)^4} + 2\,\frac{(\bvq \cdot \baq)
 (\bm{n}\cdot\baq)}{(1 - \bm{n} \cdot \bvq)^5} -\frac{(\bm{n}\cdot\baq)^2}
 {\gq^2(1 - \bm{n} \cdot \bvq)^6} \right],
 \\\label{Escal}
 \mcal{E}_{\rm scalar}&=&
 \frac{\lambda}{32 \pi^2 R^2}
 \left[\frac{\gq^2 (\bvq \cdot \baq)^2}{(1 - \bm{n} \cdot \bvq)^4}
 - 2\,\frac{(\bvq \cdot \baq)(\bm{n}\cdot\baq)}{(1 - \bm{n} \cdot \bvq)^5}
 +\frac{(\bm{n}\cdot\baq)^2}{\gq^2(1 - \bm{n} \cdot \bvq)^6} \right],
 \eeq
where we have substituted $e_{\rm eff}^2 = \lambda/2$. Adding the two
contributions the terms depending on $\bm{n} \cdot \baq$
cancel\footnote{This cancelation for the particular case of uniform
circular motion was not realized by the authors of
\cite{Athanasiou:2010pv}, because their corresponding expressions (2.20a)
and (2.20b) for $\mcal{E}_{\rm vector}$ and $\mcal{E}_{\rm scalar}$ miss
a numerical factor of 1/2 and 2 respectively.}  and we obtain a simple
result:
 \beq\label{Eradclass}
 \mcal{E}_{\rm rad}^{\rm class} =
 \frac{\lambda}{32 \pi^2}\,
 \frac{\gq^2 [\baq^2 - (\bvq \times \baq)^2]}{(\br - \brq)^2
 (1 - \bm{n} \cdot \bvq)^4}\,.
 \eeq
It is very interesting to notice that with the replacement $\lambda \to 4
\sqrt{\lambda}$, this is exactly the same as the $\Eone$ piece of the
strong coupling result in \eqn{Erad1}. Since by assumption $r \gg r_q$
and thus the retarded time can be approximated as $t_r \simeq t-r$, we
see that the only angular dependence is the boost factor $(1 - \bm{n}
\cdot \bvq)^4$ in the denominator. This means that in the
non--relativistic limit \eqn{Eradclass} becomes isotropic and reduces to
\beq\label{Eradnr}
 \mcal{E}_{\rm rad}^{\rm class} \simeq
 \frac{\lambda}{32 \pi^2}\,
 \frac{\baq^2}{r^2}.
 \eeq
Obviously this is a property which is not shared by QED, where only
vector fields are radiated, and the radiated energy as given in
\eqnum{Evec} contains anisotropic pieces.

One can compute separately the vector and scalar contributions to the
power (see again Appendix \ref{integrals} for the corresponding
integrals), which read
 \beq\label{Pvectorscalar}
 P_{\rm vector} = \frac{\lambda}{12 \pi}
 \,\gq^6\,[\baq^2 - (\bvq \times \baq)^2]\, \quad \textrm{and} \quad
 P_{\rm scalar} = \frac{\lambda}{24 \pi}
 \,\gq^6\,[\baq^2 - (\bvq \times \baq)^2]\,,
 \eeq
leading to a total power
 \beq\label{Pclass}
 P_{\rm rad}^{\rm class} = \frac{\lambda}{8 \pi}
 \,\gq^6\,[\baq^2 - (\bvq \times \baq)^2]\,,
 \eeq
which, up to the  replacement $\lambda \to 4 \sqrt{\lambda}$, is the same
as the piece $P_{\rm rad}^{(1)}$ of the corresponding supergravity
result, cf. \eqn{Prad1}.

\section{Discussion and open issues}
\label{discussion}

One of the main results of this paper is the verification of a conjecture
made in \cite{Hatta:2010dz} that for an arbitrary relativistic motion of
a heavy quark, and in the supergravity approximation to the dual string
theory, it is only the endpoint of the dual string at $z=0$ which
contributes to the radiated energy. That was first observed in
\cite{Athanasiou:2010pv} for the uniform circular motion and then
extended in \cite{Hatta:2010dz} to a general non--relativistic motion and
also, {\em mutatis mutandis}, to other types of radiation, like the decay
of a time--like wave--packet. (The dual description of the time--like
wave--packet is a supergravity field falling into AdS$_5$. Then the
corresponding statement is that the radiation is generated only from the
starting point of the trajectory at $z=0$.) This property implies that
the radiation propagates at the speed of light and therefore the
space--time distribution of this radiated energy is very similar to that
of the corresponding classical radiation, without any sign of quantum
broadening \cite{Athanasiou:2010pv,Hatta:2010dz}. As argued in
\cite{Hatta:2010dz} this leads to a radial distribution which is to
difficult to reconcile with quantum mechanics, which sheds doubts on the
validity of the supergravity approximation as the correct, dual,
description of the strong--coupling limit. Moreover, it was shown there,
for a specific example and via an admittedly heuristic calculation, that
there are particular string corrections (which in the light--cone gauge
appear as  fluctuations in the longitudinal coordinates of the string
points) which are not suppressed when $\lambda\to\infty$, and hence
should be treated as a part of the leading order theory at strong
coupling. One effect of those fluctuations (at least within the limits of
the calculation in \cite{Hatta:2010dz}) is to provide a radial broadening
for the energy distribution, in agreement with expectations from both
quantum mechanics and the UV/IR correspondence.

The proper calculation of string fluctuations in a curved space--time is
an outstanding open problem, that we shall not attempt to address here.
Rather, we would like to emphasize some curious features of the previous
results obtained in the supergravity approximation, which look rather
implausible to us on physical grounds and may represent additional
shortcomings of this approximation (besides the lack of quantum
broadening). The peculiarities to be discussed here are all associated
with the second contribution to the energy density, $\Etwo$ in
\eqn{Erad2}.

\bigskip \noindent \texttt{(i)} {\em The lack of isotropy in the
non--relativistic limit}

As mentioned in the Introduction,  at strong coupling one expects the
radiation to be isotropically distributed at large distances away from
the source \cite{HIM3,Hofman:2008ar}, except for the trivial anisotropy
introduced by the Lorentz boost. Indeed, the radiation should typically
proceed via the emission of off--shell quanta\footnote{This follows from
the uncertainty principle: quanta with a large virtuality $Q$ have a
short formation time $\Delta t\sim \omega/Q^2$, where $\omega$ is the
energy carried by the quanta. However, after being emitted, such quanta
need to further radiate to become on--shell, which explains why their
emission is suppressed at weak coupling.}, which then should evacuate
their virtuality via successive branchings. This gives rise to a partonic
cascade through which the original energy and momentum get divided among
many quanta. Due to their large number and to the absence of any
preferred pattern in the process of branching (at strong coupling), these
quanta should have an isotropic distribution. This is generally not the
case at weak coupling (say, in classical electrodynamics), although it
happens to be the case in the weak coupling limit of the ${\mathcal N}=4$
SYM theory, as shown in Sect.~\ref{classical} (see also
\cite{Hofman:2008ar}), because of the additional symmetries of this
theory.

In view of the above, we find it extremely surprising that the
supergravity result for the radiated energy density is {\em not}
isotropic in the non--relativistic limit, especially in the context of
the ${\mathcal N}=4$ SYM theory, where the isotropy is realized already
at weak coupling. Indeed, when $v_q\ll 1$, Eqs.~\eqref{Erad1} and
\eqref{Erad2} reduce to
 \beq\label{Enr1}
 \Eone \simeq \frac{\sqrt{\lambda}}{8 \pi^2}\,
 \frac{\baq^2}{r^2}\,,\eeq
 %\quad \mathrm{and} \quad
which is isotropic and similar in structure to \eqn{Eradnr}, and
respectively
 \beq\label{Enr2}
 \Etwo \simeq -\frac{\sqrt{\lambda}}{24 \pi^2}\,
 \frac{\bm{n} \cdot \dot{\bm{a}}_q -
 \left[\baq^2 -4 (\bm{n}\cdot \baq)^2
 + \bvq \cdot  \dbaq
 + (\bm{n}\cdot \bvq)(\bm{n} \cdot \dbaq)
 \right]}{r^2}.
 \eeq
which is manifestly not isotropic. Moreover, this last, anisotropic, term
can even dominate over the first one in some cases, as shown by the
example of the uniform rotation: there, $a=\omega^2 R_0=\omega v$ and
$\dot a = \omega^3 R_0=\omega^2 v$, so clearly $\dot a \gg a^2$ when
$v\ll 1$. Since moreover the sign of $\bm{n} \cdot \dot{\bm{a}}_q$ is
oscillating when changing the direction of observation, one sees that the
radiated energy density is {\em negative} in some regions, which brings
us to our second puzzle.

\bigskip \noindent \texttt{(ii)} {\em The negativity of the
radiated energy density}

In Ref.~\cite{Athanasiou:2010pv} already the authors noticed that the
energy density \eqref{Erotsum} radiated in the case of uniform rotation
can become negative in some regions of space--time. From our present
discussion we know that this behaviour must be associated with the second
piece \eqref{Erot2} of the radiation, which in some regions can become
negative and also larger in magnitude than the first piece \eqref{Erot1}.
In Ref.~\cite{Athanasiou:2010pv}, where only the relativistic case was
considered, the regions of negative energy were relatively small (and
localized near the edges of the arm of the spiral) and besides the
negative values reached by the energy in those regions were much smaller
than its positive values towards the middle of the spiral arm. In the
non--relativistic case, however, we have just seen that (for the case of
rotation at least), the second piece \eqref{Enr2} dominates over the
first one \eqref{Enr1} anywhere except at the particular points where
$\bm{n} \cdot \dbaq$ vanishes, and that the sign of this second piece
oscillates. Specifically, the non--relativistic limit of \eqref{Erotsum}
reads
 \beq\label{Erotnr}
 \mcal{E}_{\rm rad} =
 \frac{\sqrt{\lambda} \omega^2 \upsilon}{24 \pi^2 r^2}\,
 \sin \theta \sin(\phi - \omega t_r)\,,
 \eeq
which arises, as expected, from the $\bm{n} \cdot \dbaq$ term of $\Etwo$
in \eqn{Enr2}. Clearly, this (dominant) contribution to the radiated
energy density oscillates around zero with the positive and negative
maxima being of equal magnitude.

In principle, regions of negative energy density can occur in a quantum
field theory, in the process of subtraction of the ultraviolet (UV)
divergences. For the problem under consideration, such UV issues could
affect the proper energy of the heavy quark, as carried by its near
field, but on the other hand we find them rather unnatural in relation
with the far fields and the radiation. As manifest say in \eqn{Erotnr},
the length scale associated with such space--time variations in the
radiated energy is not some UV cutoff, but rather is determined by the
external force that is giving the quark the specified motion.

Note also that \eqn{Erotnr}, and more generally the  $\bm{n} \cdot \dbaq$
term of \eqn{Enr2}, do not contribute to the radiated {\em power}, since
they integrate to zero. The power appears to be dominated by the first
term \eqref{Prad1}, and thus be positive, for all the examples that we
have investigated.

To summarize, the anisotropy and the negativity of the energy density
associated with the contribution $\Etwo$ in \eqn{Erad2} look very
unnatural to us and make us feel skeptical about this particular term. In
our opinion, these unappealing features are merely an artifact of the
supergravity approximation which will be corrected after including string
fluctuations. It is also possible that this term, or at least a part of
it, represent the tail of the quark proper energy at large distances, as
suggested by the comparison between the associated power, \eqn{Prad2},
and the results in \cite{Chernicoff:2009re,Chernicoff:2009ff} (cf. the
discussion after \eqn{Eqtot}).

Also, the fact that the problems alluded to above are solely generated by
the second piece, \eqn{Erad2}, of the energy density does not mean that
we fully trust the other piece in \eqn{Erad1}. In spite of its rather
appealing structure and of its similarity with the corresponding
classical result, this term too has been produced from the string
endpoint at $z=0$ and thus it shows no quantum (radial) broadening. We
therefore believe that also this term will be modified by string
fluctuations, in the sense of acquiring a spread, but in such a way that
its spatial integral giving the power will remain unchanged. Indeed, we
believe that the correct result for the power (at least for a bounded
motion and sufficiently large times) is given by $P_{\rm rad}^{(1)}$ in
\eqn{Prad1}, because this expression has been suggested by independent
considerations (based on a world--sheet analysis) in
Refs.~\cite{Mikhailov:2003er,Chernicoff:2009re,Chernicoff:2009ff} and
because it coincides with the energy flow at the world--sheet horizon in
all the examples that have been worked out in the literature.

At this point, we should recall that Ref.~\cite{Hofman:2008ar} has
studied the angular distribution of the energy density produced by the
decay of a time--like wave--packet within AdS/CFT, and found that this is
isotropic in the supergravity approximation and it is only weakly
affected by string fluctuations (at least, in a heuristic treatment of
the latter inspired by flat--space string quantization). However, in that
particular problem there was no kinematical scale which could induce an
anisotropy (the wave--packet was spherically symmetric and at rest),
unlike in the heavy quark problem under present consideration, where
there are such scales. For us, our present results signal that, in
general, the supergravity predictions cannot be trusted neither for the
angular distribution of the radiation, nor for the radial one. Therefore,
any progress towards better understanding the effects of the string
fluctuations would be of paramount importance.

\section*{Acknowledgments}
The work of Y.~H. is supported by Special Coordination Funds for
Promoting Science and Technology of the Ministry of Education, the
Japanese Government. The work of E.~I. is supported in part by Agence
Nationale de la Recherche via the programme ANR-06-BLAN-0285-01. The work
of A.H.~M. is supported in part by the US Department of Energy.

\appendix

\section{Useful integrals}
\label{integrals}

Here we list some standard integrals which are useful when calculating
the total power. With $\bm{\upsilon}$ the velocity, $\gamma$ the Lorentz
boost factor, $\bm{a}$ an arbitrary vector and $\bm{n}$ the unit vector
 \beq\label{bmn}
 \bm{n} = (\sin\theta\cos\phi,\sin\theta\sin\phi,\cos\theta),
 \eeq
we have
 \beq\label{inta0d3}
 \int \dif \Omega\,\frac{1}{(1 - \bm{n}\cdot\bm{v})^3} &=&
 4 \pi \gamma^4,
 \\ \label{inta1d4}
 \int \dif \Omega\,\frac{\bm{n}\cdot\bm{a}}{(1 - \bm{n}\cdot\bm{v})^4} &=&
 \frac{16 \pi}{3}\,\gamma^6\, \bm{\upsilon} \cdot \bm{a},
 \\ \label{inta2d5}
 \int \dif \Omega\,\frac{(\bm{n}\cdot\bm{a})^2}
 {(1 - \bm{n}\cdot\bm{v})^5}&=&
 \frac{4 \pi}{3}\,\gamma^6 a^2 +
 8 \pi \gamma^8 (\bm{\upsilon} \cdot \bm{a}) ^2,
 \eeq
and also
 \beq\label{intA}
 \int \dif \Omega\,\frac{1 + \bm{n}\cdot\bm{v}}
 {(1 - \bm{n}\cdot\bm{v})^2} &=&
 8 \pi \gamma^2 -
 \frac{4 \pi}{\upsilon}\, \tanh^{-1}\upsilon,
 \\ \label{intB}
 \int \dif \Omega\,
 \frac{(\bm{n} \times \bm{\upsilon})^2}{(1 - \bm{n}\cdot \bm{v})^3}
 &=& 4 \pi \gamma^2 -
 \frac{4 \pi}{\upsilon}\, \tanh^{-1}\upsilon.
\eeq
An easy way to perform all the above or similar integrations is to
assume, without any loss of generality, that instantaneously the particle
is moving along the third axis, that is $\bm{\upsilon} = (0,0,\upsilon)$.
Then it is straightforward to perform the integral $\int \dif \Omega\,
[(\bm{n}\cdot\bm{a})^p/(\alpha - \upsilon\cos\theta)]$ with integer $p\ge
0$ and arbitrary $\alpha > \upsilon$. The desired integrals follow by an
appropriate number of differentiations with respect to $\alpha$ evaluated
at $\alpha=1$.

%\bibliographystyle{JHEP}
%\bibliography{ADSref}

\providecommand{\href}[2]{#2}\begingroup\raggedright\endgroup

\end{document}